\documentclass[aps,preprint]{revtex4}
\usepackage{eurosym}
\usepackage{amsfonts}
\usepackage{amsmath}
\usepackage{amssymb}

\begin{document}

\title{Exact Kerr--Schild spacetimes from linearized kinetic gravity braiding%
}
\author{Bence Juh\'{a}sz$^{1}$}
\author{L\'{a}szl\'{o} \'{A}rp\'{a}d Gergely$^{1,2}$}
\affiliation{$^{1}$Department of Theoretical Physics, University of Szeged, Tisza Lajos
krt. 84-86, H-6720 Szeged, Hungary}
\affiliation{\thinspace $^{2}$Department of Theoretical Physics, HUN-REN Wigner Research
Centre for Physics, Konkoly-Thege Mikl\'{o}s \'{u}t 29-33, H-1121 Budapest,
Hungary}

\begin{abstract}
We generalize our recent work on k-essence sourcing Kerr--Schild spacetimes
to kinetic gravity braiding scalar field. For k-essence, in order a
perturbative Kerr--Schild type solution to become exact, the k-essence
Lagrangian was either linear in the kinetic term (with the Kerr--Schild
congruence autoparallel) or unrestricted, provided the scalar gradient along
the congruence vanishes. A similar reasoning for the pure kinetic braiding
contribution leads to either a vanishing Lagrangian or a scalar which is
constant along the congruence. From the scalar dynamics we also derive an
accompanying constraint. Finally, we discuss pp-waves, an example of
Kerr--Schild spacetime generated by a constant k-essence along the
Kerr--Schild congruence with vanishing Lagrangian. This allows for the
construction of a Fock-type space, consisting of a tower of Kerr--Schild
maps first yielding a vacuum pp-wave from flat spacetime; next a k-essence
generated pp-wave from the vacuum pp-wave; and finally an arbitrary number
of k-essence pp-waves with different, retarded time dependent metric
functions.
\end{abstract}

\date{\today }
\maketitle

\section{Introduction}

Kerr--Schild maps $\tilde{g}_{ab}=g_{ab}+\lambda l_{a}l_{b}$ with $\lambda $
an arbitrary parameter and $l_{a}l^{a}=0$ often appear as recipes to
generate exact solutions from known ones through a null congruence in such a
way that only the generator of the light cone along the congruence stays
unaffected \cite{KerrSchild}. Plane-fronted waves with parallel propagation
(pp-waves) are generated from a flat seed metric $g_{ab}=\eta _{ab}$ through
a Kerr--Schild map with a null congruence with all optical scalars
(expansion, twist, shear) vanishing. Schwarzschild black hole metrics emerge
in a similar fashion, this time the null congruence having expansion. The
Kerr--Schild map on a flat seed spacetime with a particular null congruence,
which has both expansion and twist, generates Kerr black hole metrics. In
all these cases the null congruence stays shearfree. Shearing Kerr--Schild
congruences generate one of the K\'{o}ta--Perj\'{e}s metrics or its
nontwisting limit, the Kasner metric from a type N vacuum Kundt metric or a
vacuum pp-wave, respectively \cite{GALPZ,GALPZ2,GALPZ3}. Recently,
Kerr--Schild spacetimes generated attention through attempts to relate gauge
and gravitational theories \cite{KentZimmerman}.

Kerr--Schild vacuum spacetimes emerging perturbatively for small $\lambda $
have the remarkable property that they become exact solutions of the
Einstein equations, holding for arbitrary values of the Kerr--Schild
parameter, as proved by Xanthopoulos \cite{Xanthopoulos}. As an extension of
this result for the nonvacuum case, the pair $\left( \tilde{g}%
_{ab},T_{ab}+\lambda T_{ab}^{\left( 1\right) }\right) $ solving the
linearized equation generates the exact solution $\left( \tilde{g}%
_{ab},T_{ab}+\lambda T_{ab}^{\left( 1\right) }+\lambda
^{2}l_{(a}T_{b)c}^{\left( 1\right) }l^{c}\right) $, provided the null
congruence is autoparallel (otherwise, a similar, more involved result
holds) \cite{GAL}.

Introducing at least one additional degree of freedom,\ complementing the
tensorial ones of general relativity could provide geometric explanations
for dark matter or dark energy, also to model inflation or low energy
modifications due to quantum gravity. The kinetic gravity braiding class of
modified gravity theories \cite{KGB} incorporates a scalar field into the
gravitational sector, but still allows for the propagation of the tensorial
modes at all frequencies with the speed of light in vacuum and is consistent
with all available observations. The dependence of the Lagrangian on the
scalar $\phi $ of such theories is only through $\phi ,~\square \phi $ and
the kinetic term $X=-\left( g^{ab}\nabla _{a}\phi \nabla _{b}\phi \right)
/2\ $(with $g^{ab}$ the inverse metric, $\nabla _{a}$ the Levi-Civita
connection and $\square =\nabla _{a}\nabla ^{a}$). We further restrict this
class by assuming minimal coupling of the kinetic gravity braiding scalar
field, supressing any time evolution of the gravitational constant. This
will render the equation of motion into the form of an Einstein equation,
with the left-hand side the Einstein tensor and the right-hand side
containing the contributions of the scalar.

This is a subclass of the Horndeski theories \cite{Horndeski1,Deffeayet},
which have the convenient property that both the metric and the scalar
evolve through second order dynamics (thus no Ostrogradsky instabilities
occur). In addition, the propagation speed of the tensorial modes (the
gravitational waves) is the speed of light in vacuum at all frequencies \cite%
{GW,GW0,GW1,GW2,GW3} (thus they comply with observations of high frequency
gravitational waves by LIGO and Virgo \cite{GW}).

Cosmological evolutions in kinetic gravity braiding theories were discussed
in Refs. \cite{bigrip,phantom}, which identified evolutions leading into one
of the following cases: de Sitter state, the future Big Rip singularity
occurring in finite time or diverging energy density.

In a previous paper \cite{kessence} we have analyzed Kerr--Schild maps for
k-essence spacetimes whose Lagrangians do not involve $\square \phi $,
identifying the Lagrangians linear in $X$ as able to reproduce the property
that perturbative Kerr--Schild solutions are also exact (for autoparallel
Kerr--Schild congruence). In this paper we extend these investigations both
to special cases of k-essence and to the more general case, which includes $%
\square \phi $.

In Section 2 we present the kinetic gravity braiding dynamics and the
existing results on perturbative Kerr--Schild spacetimes, which generate
exact solutions in the presence of matter sources, specifying for the case
of k-essence. We then turn to pure kinetic gravity scalar fields in Section
3, discussing the requirements for the exactness of perturbative solutions
and we derive a scalar constraint in Section 4. In Section 5 we investigate
pp-waves generated by k-essence. We summarize our results in the Section 6.

\section{Preliminaries}

\subsection{Kinetic gravity braiding}

The action for the minimally coupled kinetic gravity is a sum of the
Einstein--Hilbert action with the scalar contribution%
\begin{eqnarray}
S_{\phi } &=&\int \mathrm{d}^{4}x\mathfrak{L}_{\phi }~,  \notag \\
\mathfrak{L}_{\phi } &=&\sqrt{-\mathfrak{g}}\left[ \mathit{L}_{2}\left( \phi
,X\right) +G_{3}(\phi ,X)\square \phi \right] ~,  \label{action}
\end{eqnarray}%
where $\mathit{L}_{2}$ is the Lagrangian of k-essence and $G_{3}(\phi
,X)\square \phi $ is the pure kinetic gravity braiding contribution. Both $%
\mathit{L}_{2}$ and $G_{3}$ are arbitrary functions of $\phi $ and the
kinetic term $X$. It is often convenient to rewrite the scalar action
through a partial integration into the form containing the Lagrangian density%
\begin{equation}
\mathfrak{L}_{\phi }^{\text{KGB}}=\sqrt{-\mathfrak{g}}\left[ F\left( \phi
,X\right) +H\left( \phi ,X\right) \nabla _{c}X\nabla ^{c}\phi \right] ~,
\label{actionS1}
\end{equation}%
with $F=\mathit{L}_{2}+2G_{3\phi }X$ and $H=-G_{3X}$.

Variation with respect to the inverse metric $g^{ab}$ through the
prescription 
\begin{equation}
T_{ab}=\dfrac{-2}{\sqrt{-\mathfrak{g}}}\dfrac{\delta S}{\delta g^{ab}}\ 
\label{EM}
\end{equation}%
gives the energy-momentum tensor 
\begin{eqnarray}
T_{ab}^{\phi } &=&T_{ab}^{F}+T_{ab}^{H}~,  \label{enmom} \\
T_{ab}^{F} &=&F_{X}\nabla _{a}\phi \nabla _{b}\phi +g_{ab}F~,  \label{T2} \\
T_{ab}^{H} &=&H\left( g_{ab}\nabla _{c}X\nabla ^{c}\phi -2\nabla
_{(a}X\nabla _{b)}\phi \right)  \notag \\
&&+\left( 2H_{\phi }X-H\square \phi \right) \nabla _{a}\phi \nabla _{b}\phi
~.  \label{T3}
\end{eqnarray}%
(Derivatives with respect to $\phi $ and $X$ are denoted by the respective
subscripts.) Variation with respect to the scalar gives the scalar equation
of motion $\left( -\mathfrak{g}\right) ^{-1/2}\delta S_{\phi }/\delta \phi
=0 $ as \cite{KGBFluid}:%
\begin{eqnarray}
&&F_{\phi }-2F_{\phi X}X-F_{XX}\nabla _{a}\phi \nabla _{b}\phi \left( \nabla
^{a}\nabla ^{b}\phi \right) +F_{X}\square \phi  \notag \\
&&+H\left[ R_{ab}\nabla ^{a}\phi \nabla ^{b}\phi +\left( \nabla _{a}\nabla
_{b}\phi \right) \left( \nabla ^{a}\nabla ^{b}\phi \right) -\left( \square
\phi \right) ^{2}\right]  \notag \\
&&-2H_{\phi }\left[ \nabla _{a}\phi \nabla _{b}\phi \left( \nabla ^{a}\nabla
^{b}\phi \right) -2X\square \phi \right]  \notag \\
&&+H_{X}\nabla _{a}\phi \nabla _{b}\phi \left[ \left( \nabla ^{a}\nabla
^{b}\phi \right) \square \phi -\left( \nabla ^{i}\nabla ^{b}\phi \right)
\left( \nabla _{i}\nabla ^{a}\phi \right) \right]  \notag \\
&&-4H_{\phi \phi }X^{2}-2H_{\phi X}X\nabla _{a}\phi \nabla _{b}\phi \left(
\nabla ^{a}\nabla ^{b}\phi \right) =0~.  \label{scalarEq}
\end{eqnarray}%
In deriving this result, the derivative of $X$ was replaced by 
\begin{equation}
\nabla _{a}X=-\nabla ^{c}\phi \nabla _{a}\nabla _{c}\phi ~.  \label{nablatX}
\end{equation}

Variation of the Einstein--Hilbert action with respect to the inverse metric
complements the set of dynamical equations with the Einstein equation%
\begin{equation}
G_{ab}=T_{ab}~.
\end{equation}%
Here the choice of units $8\pi G=c=1$ has been implemented. From here, the
Ricci tensor appearing in the scalar equation (\ref{scalarEq}) of motion can
be expressed in terms of the scalar energy-momentum tensor (\ref{enmom}) and
its trace%
\begin{equation}
T^{\phi }=T^{F}+T^{H}~,
\end{equation}%
with%
\begin{eqnarray}
T^{F} &=&-2XF_{X}+4F~, \\
T^{H} &=&2H\nabla _{c}X\nabla ^{c}\phi -4H_{\phi }X^{2}+2XH\square \phi ~
\end{eqnarray}%
as

\begin{equation}
R_{ab}=T_{ab}^{\phi }-\frac{1}{2}g_{ab}T^{\phi }~.
\end{equation}%
With this the curvature term can be eliminated from the equation of motion
of the scalar field. By also employing Eq. (\ref{nablatX}) it exhibits the
structure%
\begin{equation}
\emph{Sc}\left[ F\right] +\emph{Sc}\left[ H\right] +\emph{Sc}\left[ H^{2}%
\right] +\emph{Sc}\left[ FH\right] =0~,
\end{equation}%
with the contributions%
\begin{align}
& \emph{Sc}\left[ F\right] =F_{\phi }-2F_{\phi X}X-F_{XX}\nabla _{a}\phi
\nabla _{b}\phi \left( \nabla ^{a}\nabla ^{b}\phi \right) +F_{X}\square \phi
~, \\
& \emph{Sc}\left[ H\right] =H\left[ \left( \nabla _{a}\nabla _{b}\phi
\right) \left( \nabla ^{a}\nabla ^{b}\phi \right) -\left( \square \phi
\right) ^{2}\right]  \notag \\
& -2H_{\phi }\left[ \nabla _{a}\phi \nabla _{b}\phi \left( \nabla ^{a}\nabla
^{b}\phi \right) -2X\square \phi \right]  \notag \\
& +H_{X}\nabla _{a}\phi \nabla _{b}\phi \left[ \left( \nabla ^{a}\nabla
^{b}\phi \right) \square \phi -\left( \nabla ^{i}\nabla ^{b}\phi \right)
\left( \nabla _{i}\nabla ^{a}\phi \right) \right]  \notag \\
& -4H_{\phi \phi }X^{2}-2H_{\phi X}X\nabla _{a}\phi \nabla _{b}\phi \left(
\nabla ^{a}\nabla ^{b}\phi \right) ~, \\
& \emph{Sc}\left[ H^{2}\right] =2H^{2}\left( -2\nabla ^{a}\phi \nabla
^{b}\phi \nabla _{a}\nabla _{b}\phi -X\square \phi \right) X+4HH_{\phi
}X^{3}~, \\
& \emph{Sc}\left[ FH\right] =2H\left( F+F_{X}X\right) X~.
\end{align}%
The first of these characterizes pure k-essence (only $F$-terms), the second
and third pure kinetic gravity braiding scalar (with both linear and
nonlinear contributions in $H$ and its derivatives), while the last one is
an interaction term emerging from the combination of both types of
contributions to the action.

\subsection{Kerr--Schild maps}

In Ref. \cite{GAL}, it was shown that the Kerr--Schild transformed Ricci
tensor%
\begin{equation}
\tilde{R}_{ab}=R_{ab}+\lambda R_{ab}^{(1)}+\lambda ^{2}R_{ab}^{(2)}+\lambda
^{3}R_{ab}^{(3)}~,
\end{equation}%
is a third order polynomial in $\lambda $, with the contributions%
\begin{eqnarray}
R_{ab}^{\left( 1\right) } &=&\nabla _{c}\left[ \nabla _{(a}\left(
l_{b)}l^{c}\right) -\frac{1}{2}\nabla ^{c}\left( l_{a}l_{b}\right) \right] ~,
\label{Ric1} \\
R_{ab}^{\left( 2\right) } &=&\nabla _{c}l^{c}l_{(a}Dl_{b)}+\frac{1}{2}%
Dl_{a}Dl_{b}+l_{(a}DDl_{b)}+l_{a}l_{b}\nabla _{c}l_{d}\nabla ^{\lbrack
c}l^{d]}-\left( Dl^{c}\right) \nabla _{c}l_{(a}l_{b)}~,  \label{Ric2} \\
R_{ab}^{\left( 3\right) } &=&-\frac{1}{2}l_{a}l_{b}Dl^{c}Dl_{c}~.
\label{Ric3}
\end{eqnarray}%
By taking the Kerr--Schild transformed energy-momentum tensor as an infinite
series in $\lambda $%
\begin{equation}
\tilde{T}_{ab}=\sum_{k=0}^{\infty }\lambda ^{k}T_{ab}^{(k)}~,  \label{Texp}
\end{equation}%
the Einstein equations restrict all $k>3$ contributions to vanish, $%
T_{ab}^{(k>3)}=0$. If such contributions were to exist in the
energy-momentum tensor, it would not be able to source Kerr--Schild metrics.

With $\lambda $ arbitrary, the Einstein equation decouples into the
following $\lambda $, $\lambda ^{2}$ and $\lambda ^{3}$ contributions:%
\begin{equation}
R_{ab}^{\left( 1\right) }=T_{ab}^{(1)}+\frac{1}{2}g_{ab}\left(
T_{cd}l^{c}l^{d}-T^{\left( 1\right) }\right) -\frac{1}{2}l_{a}l_{b}T~,
\label{E1}
\end{equation}%
\begin{equation}
R_{ab}^{\left( 2\right) }=T_{ab}^{(2)}+\frac{1}{2}g_{ab}\left(
T_{cd}^{\left( 1\right) }l^{c}l^{d}-T^{\left( 2\right) }\right) +\frac{1}{2}%
l_{a}l_{b}\left( T_{cd}l^{c}l^{d}-T^{\left( 1\right) }\right) ~,  \label{E2}
\end{equation}%
\begin{equation}
R_{ab}^{\left( 3\right) }=T_{ab}^{(3)}+\frac{1}{2}g_{ab}\left(
T_{cd}^{\left( 2\right) }l^{c}l^{d}-T^{\left( 3\right) }\right) +\frac{1}{2}%
l_{a}l_{b}\left( T_{cd}^{\left( 1\right) }l^{c}l^{d}-T^{\left( 2\right)
}\right) ~.  \label{E3}
\end{equation}%
In Ref. \cite{GAL} conditions on the energy-momentum tensor were identified,
which assure that any solution of Eq. (\ref{E1}) also satisfies Eqs. (\ref%
{E2})-(\ref{E3}). These were:%
\begin{eqnarray}
T_{ab}^{\left( 3\right) } &=&-\frac{3}{4}l_{a}l_{b}\left(
Dl^{c}Dl_{c}\right) ~,  \label{cond3} \\
2T_{ab}^{\left( 2\right) } &=&2l_{(a}T_{b)c}^{\left( 1\right) }l^{c}-\frac{1%
}{2}g_{ab}\left( Dl^{c}Dl_{c}\right) +Dl_{a}Dl_{b}-l_{a}l_{b}\left( \nabla
_{c}Dl^{c}\right)  \notag \\
&&+l_{(a}\left[ Dl_{b)}\left( \nabla _{c}l^{c}\right) +DDl_{b)}+\left(
\nabla _{b)}l_{c}-2\nabla _{\left\vert c\right\vert }l_{b)}\right) Dl^{c}%
\right] ~.  \label{cond21}
\end{eqnarray}%
For an autoparallel congruence $l^{a}$ (thus $Dl^{a}\propto l^{a}$, with $%
D=l^{b}\nabla _{b}$ the covariant derivative along $l^{a}$), these
conditions further simplified, requiring that the third-order contribution
to the Kerr--Schild-transformed energy-momentum tensor vanishes, while the
second- and first-order contributions obey a constraint: 
\begin{equation}
T_{ab}^{(3)}=0~,\quad T_{ab}^{(2)}=l_{(a}T_{b)c}^{\left( 1\right) }l^{c}~.
\label{autopcond}
\end{equation}

Perturbative solutions emerge from Eq. (\ref{E1}). Whenever the above
conditions hold for autoparallel Kerr--Schild congruences, these solutions
also generate the exact solution.

\subsection{K-essence under Kerr--Schild maps}

The Kerr--Schild transformation does not affect $\phi $ or $\nabla _{c}\phi $%
,\ however it changes $g_{ab}$, $X$, $\nabla _{c}X$ and $\square \phi $.
From among these variables, only $g_{ab}$ and $X$ appear in the k-essence
part $T_{ab}^{F}$. As the inverse metric transforms according to $\tilde{g}%
^{ab}=g^{ab}-\lambda l^{a}l^{b}$, the kinetic variable $X$ changes into%
\begin{equation}
\tilde{X}=X+\frac{\lambda }{2}\left( D\phi \right) ^{2}~.  \label{Xtil}
\end{equation}%
The Kerr--Schild transformation in the k-essence case was analyzed in detail
in our previous work \cite{kessence}. As a main result, the functional form
of the free function $F$\ characterizing the k-essence was restricted to be
linear in $X$ by the requirement to make perturbative solutions generating
the exact ones.

Any Kerr--Schild null congruence still needs to obey the linearized equation
(\ref{E1}), with the first-order contribution 
\begin{equation}
T_{ab}^{F(1)}=l_{a}l_{b}F+\frac{1}{2}\left( F_{X^{2}}\nabla _{a}\phi \nabla
_{b}\phi +g_{ab}F_{X}\right) \left( D\phi \right) ^{2},  \label{T1kessence}
\end{equation}%
calculated in Ref. \cite{kessence}. For the allowed 
\begin{equation}
F=C\left( \phi \right) X-U\left( \phi \right) ~,  \label{Flin}
\end{equation}%
linear in $X$, with $C$ and $U$ two arbitrary functions of $\phi $, energy
momentum tensor (\ref{T2}), the linearized equation (\ref{E1}) reads 
\begin{equation}
\nabla _{c}\left[ \nabla _{(a}\left( l_{b)}l^{c}\right) -\frac{1}{2}\nabla
^{c}\left( l_{a}l_{b}\right) \right] =l_{a}l_{b}U~.  \label{linKSkessence}
\end{equation}

\section{Pure kinetic gravity braiding under Kerr--Schild maps}

In this section we discuss the changes occurring under Kerr--Schild maps in
the pure kinetic gravity braiding contribution $T_{ab}^{H}$. For this,
employing Eq. (\ref{Xtil}) we calculate 
\begin{eqnarray}
\tilde{\nabla}_{c}\tilde{X} &=&\nabla _{c}X+\lambda D\phi \nabla _{c}D\phi 
\notag \\
&=&\nabla _{c}X+\lambda \left[ \left( \nabla _{c}l^{a}\right) \nabla
_{a}\phi +l^{a}\nabla _{c}\nabla _{a}\phi \right] D\phi ~.  \label{nablaXKS}
\end{eqnarray}%
Next, we need the Kerr--Schild transformation of the Christoffel symbols,
emerging as%
\begin{equation}
\tilde{\Gamma}_{ab}^{i}=\Gamma _{ab}^{i}+\lambda g^{ij}\left[ \nabla
_{(a}\left( l_{j}l_{b)}\right) -\frac{1}{2}\nabla _{j}\left(
l_{a}l_{b}\right) \right] +\frac{\lambda ^{2}}{2}l^{i}D\left(
l_{a}l_{b}\right) ~,
\end{equation}%
to obtain 
\begin{equation}
\tilde{\nabla}_{a}\tilde{\nabla}_{b}\phi =\nabla _{a}\nabla _{b}\phi
-\lambda \left[ \nabla _{(a}\left( l_{c}l_{b)}\right) -\frac{1}{2}\nabla
_{c}\left( l_{a}l_{b}\right) \right] g^{cd}\nabla _{d}\phi -\frac{\lambda
^{2}}{2}D\left( l_{a}l_{b}\right) D\phi ~,
\end{equation}%
which yields the Kerr--Schild transformation of the d'Alembertian:%
\begin{eqnarray}
\tilde{\square}\phi &=&\left( g^{ab}-\lambda l^{a}l^{b}\right) \tilde{\nabla}%
_{a}\tilde{\nabla}_{b}\phi  \notag \\
&=&\square \phi -\lambda g^{ab}g^{cd}\nabla _{a}\left( l_{c}l_{b}\right)
\nabla _{d}\phi -\lambda l^{a}l^{b}\nabla _{a}\nabla _{b}\phi  \notag \\
&=&\square \phi -\lambda \left[ D^{2}\phi +\left( \nabla _{a}l^{a}\right)
D\phi \right] ~.  \label{dalKS}
\end{eqnarray}%
The kinetic gravity braiding energy-momentum tensor under the Kerr--Schild
map formally transforms into a second order polynomial in $\lambda $: 
\begin{eqnarray}
\tilde{T}_{ab}^{H} &=&H(\phi ,\tilde{X})g_{ab}\nabla _{c}X\nabla ^{c}\phi
-2H(\phi ,\tilde{X})\nabla _{(a}X\nabla _{b)}\phi  \notag \\
&&+2H_{\phi }(\phi ,\tilde{X})X\nabla _{a}\phi \nabla _{b}\phi -H(\phi ,%
\tilde{X})\square \phi \nabla _{a}\phi \nabla _{b}\phi ~  \notag \\
&&+\lambda H(\phi ,\tilde{X})l_{a}l_{b}\nabla _{c}X\nabla ^{c}\phi +\lambda
H(\phi ,\tilde{X})g_{ab}D\phi \nabla _{c}D\phi \nabla ^{c}\phi  \notag \\
&&-2\lambda H(\phi ,\tilde{X})D\phi \nabla _{(a}D\phi \nabla _{b)}\phi
+\lambda H_{\phi }(\phi ,\tilde{X})\left( D\phi \right) ^{2}\nabla _{a}\phi
\nabla _{b}\phi  \notag \\
&&+\lambda H(\phi ,\tilde{X})\left( \left[ D^{2}\phi +\left( \nabla
_{c}l^{c}\right) D\phi \right] \right) \nabla _{a}\phi \nabla _{b}\phi ~ 
\notag \\
&&+\lambda ^{2}H(\phi ,\tilde{X})l_{a}l_{b}D\phi \left( \nabla _{c}D\phi
\right) \nabla ^{c}\phi ~.  \label{TtildeH}
\end{eqnarray}%
In what follows, we will implement the additional expansion of $\tilde{X}$.

\subsection{Infinitesimal Kerr--Schild maps}

The kinetic gravity braiding contribution to the energy-momentum tensor can
be fully expanded in powers of a small $\lambda $. For this we insert
infinite power serie expansions of the functions with argument $\tilde{X}$:%
\begin{eqnarray}
H(\phi ,\tilde{X}) &=&\sum_{j=0}^{\infty }\left( \frac{\lambda }{2}\right)
^{j}\frac{\left( D\phi \right) ^{2j}}{j!}H_{X^{j}}(\phi ,X)~,  \label{Hexp}
\\
H_{\phi }(\phi ,\tilde{X}) &=&\sum_{j=0}^{\infty }\left( \frac{\lambda }{2}%
\right) ^{j}\frac{\left( D\phi \right) ^{2j}}{j!}H_{\phi X^{j}}(\phi ,X)~.
\end{eqnarray}%
The transformed energy-momentum tensor contains the contribution of the
original kinetic gravity braiding field,%
\begin{equation}
T_{ab}^{(0)}=T_{ab}^{H}~,
\end{equation}%
together with the leading order correction%
\begin{eqnarray}
T_{ab}^{(1)} &=&\frac{\left( D\phi \right) ^{2}}{2}\left[ g_{ab}\nabla
_{c}X\nabla ^{c}\phi -2\nabla _{(a}X\nabla _{b)}\phi -\nabla _{a}\phi \nabla
_{b}\phi \left( \square \phi -2X\partial _{\phi }\right) \right] H_{X} 
\notag \\
&&+\left[ l_{a}l_{b}\nabla _{c}X\nabla ^{c}\phi +g_{ab}D\phi \left( \nabla
_{c}D\phi \right) \nabla ^{c}\phi -2D\phi \nabla _{(a}D\phi \nabla _{b)}\phi %
\right] H  \notag \\
&&~+\nabla _{a}\phi \nabla _{b}\phi \left[ D^{2}\phi +\left( \nabla
_{c}l^{c}\right) D\phi +\left( D\phi \right) ^{2}\partial _{\phi }\right] H~,
\label{T1KGB}
\end{eqnarray}%
and higher order ($k\geq 2$) contributions%
\begin{eqnarray}
T_{ab}^{(k)} &=&\frac{\left( D\phi \right) ^{2k}}{2^{k}k!}\left[
g_{ab}\nabla _{c}X\nabla ^{c}\phi -2\nabla _{(a}X\nabla _{b)}\phi -\square
\phi \nabla _{a}\phi \nabla _{b}\phi +2X\nabla _{a}\phi \nabla _{b}\phi
\partial _{\phi }\right] H_{X^{k}}  \notag \\
&&+\frac{\left( D\phi \right) ^{2\left( k-1\right) }}{2^{k-1}\left(
k-1\right) !}\left\{ l_{a}l_{b}\nabla _{c}X\nabla ^{c}\phi +g_{ab}D\phi
\left( \nabla _{c}D\phi \right) \nabla ^{c}\phi -2D\phi \nabla _{(a}D\phi
\nabla _{b)}\phi \right.  \notag \\
&&+\left. \left[ D^{2}\phi +\left( \nabla _{c}l^{c}\right) D\phi \right]
\nabla _{a}\phi \nabla _{b}\phi +\left( D\phi \right) ^{2}\nabla _{a}\phi
\nabla _{b}\phi \partial _{\phi }\right\} H_{X^{k-1}}  \notag \\
&&+\frac{\left( D\phi \right) ^{2k-3}}{2^{k-2}\left( k-2\right) !}%
l_{a}l_{b}\left( \nabla _{c}D\phi \right) \nabla ^{c}\phi H_{X^{k-2}}~.
\label{TkKGB}
\end{eqnarray}%
These are the coefficients appearing in the final polynomial expansion in $%
\lambda $.

\subsection{Kinetic gravity braiding with $H$ linear in $X$}

A condition necessary to have a Kerr--Schild type spacetime is $%
T_{ab}^{(k)}=0$, $k\geq 4$, therefore we announce

\emph{Theorem 1. }Kinetic gravity braiding scalar fields with $H$ linear in $%
X$ (or quadratic in $X$, when $\nabla _{c}\phi \left( \nabla ^{c}D\phi
\right) =0$\ also holds) generate at most third order energy-momentum
tensors in $\lambda $ under infinitesimal Kerr--Schild maps.

\emph{Proof. }The fourth order energy-momentum tensor contribution reads%
\begin{eqnarray}
T_{ab}^{(4)} &=&\frac{1}{16}\frac{\left( D\phi \right) ^{8}}{4!}\left[
g_{ab}\nabla _{c}X\nabla ^{c}\phi -2\nabla _{(a}X\nabla _{b)}\phi -\square
\phi \nabla _{a}\phi \nabla _{b}\phi +2X\nabla _{a}\phi \nabla _{b}\phi
\partial _{\phi }\right] H_{X^{4}}  \notag \\
&&+\frac{1}{8}\frac{\left( D\phi \right) ^{6}}{3!}\left[ l_{a}l_{b}\nabla
_{c}X\nabla ^{c}\phi +g_{ab}D\phi \left( \nabla _{c}D\phi \right) \nabla
^{c}\phi -2D\phi \nabla _{(a}D\phi \nabla _{b)}\phi \right] H_{X^{3}}  \notag
\\
&&+\frac{1}{8}\frac{\left( D\phi \right) ^{6}}{3!}\left[ \left[ D^{2}\phi
+\left( \nabla _{c}l^{c}\right) D\phi \right] \nabla _{a}\phi \nabla
_{b}\phi +\left( D\phi \right) ^{2}\nabla _{a}\phi \nabla _{b}\phi \partial
_{\phi }\right] H_{X^{3}}  \notag \\
&&+\frac{1}{4}\frac{\left( D\phi \right) ^{5}}{2!}l_{a}l_{b}\left( \nabla
_{c}D\phi \right) \nabla ^{c}\phi H_{X^{2}}~.  \label{4}
\end{eqnarray}%
Assuming $T_{ab}^{(4)}=0$ and contracting with $l^{b}$ gives%
\begin{eqnarray}
0 &=&\frac{\left( D\phi \right) ^{2}}{8}\left[ l_{a}\nabla _{c}X\nabla
^{c}\phi -DX\nabla _{a}\phi -D\phi \left( \nabla _{a}X+\square \phi \nabla
_{a}\phi -2X\nabla _{a}\phi \partial _{\phi }\right) \right] H_{X^{4}} 
\notag \\
&&+\left[ l_{a}D\phi \left( \nabla _{c}D\phi \right) \nabla ^{c}\phi -\left(
D\phi \right) ^{2}\left( \nabla _{a}D\phi -\left( \nabla _{c}l^{c}\right)
\nabla _{a}\phi \right) +\left( D\phi \right) ^{3}\nabla _{a}\phi \partial
_{\phi }\right] H_{X^{3}}~.  \label{4l}
\end{eqnarray}%
Further contracting with $l^{a}$ yields%
\begin{eqnarray}
0 &=&\frac{D\phi }{8}\left[ 2DX+D\phi \left( \square \phi -2X\partial _{\phi
}\right) \right] H_{X^{4}}  \notag \\
&&+\left[ D^{2}\phi -D\phi \left( \nabla _{c}l^{c}\right) -\left( D\phi
\right) ^{2}\partial _{\phi }\right] H_{X^{3}}~.  \label{4ll}
\end{eqnarray}%
The simplest solution to this equation is provided by the quadratic
expression $H=A\left( \phi \right) X^{2}+B\left( \phi \right) X-V\left( \phi
\right) $, with $A,B,V$ arbitrary functions of the scalar. Reinserting this
into Eq. (\ref{4}), the condition $T_{ab}^{(4)}=0$ leads to%
\begin{equation}
0=\nabla _{c}\phi \left( \nabla ^{c}D\phi \right) H_{X^{2}}~,
\end{equation}%
which vanishes either for a function 
\begin{equation}
H=B\left( \phi \right) X-V\left( \phi \right) ~,  \label{Hlin}
\end{equation}%
linear in $X$, or for a particular scalar field $\nabla _{c}\phi \left(
\nabla ^{c}D\phi \right) =0$. Either of these conditions further implies the
vanishing of all higher order contributions to $\tilde{T}_{ab}^{H}$.

Targeting generic classes of the scalar field, we adopt the linear function (%
\ref{Hlin}). While this obeys the conditions $T_{ab}^{\left( k\geq 4\right)
} $ necessary to source Kerr--Schild spacetimes with infinitesimal parameter 
$\lambda $, it could equally lead to Kerr--Schild spacetimes with arbitrary
parameter $\lambda $, as can be seen from the identical expansions $H\left(
\phi ,\tilde{X}\right) =BX-V+\lambda \left( D\phi \right) ^{2}B/2$ obtained
from Eq. (\ref{Xtil}) and from Eq. (\ref{Hexp}) specified for the linear $H$.

\subsection{Perturbative Kerr--Schild solutions are not exact}

Next, we announce the following result holding for pure kinetic gravity
braiding theories.

\emph{Theorem 2.} For autoparallel Kerr--Schild null congruence and $H$
linear in $X$ the perturbative solutions can be exact only if the Lagrangian
of the pure kinetic braiding scalar field vanishes.

\emph{Proof: }For autoparallel null congruences $Dl^{a}\propto l^{a}$ the
condition $T_{ab}^{\left( 3\right) }=0$ should be imposed in order to get
Kerr--Schild metrics. A glance on Eq. (\ref{TkKGB}) immediately gives
(unless $D\phi =0$) 
\begin{equation}
0=\nabla _{c}\phi \left( \nabla ^{c}D\phi \right) B~.
\end{equation}%
Unless $\nabla _{c}\phi \left( \nabla ^{c}D\phi \right) =0$, the condition
becomes $B=0$, hence the function $H=-V\left( \phi \right) $ loses any $X$%
-dependence.

The expressions%
\begin{equation}
T_{bc}^{(1)}l^{c}=-\left[ l_{b}D\phi \left( \nabla _{i}D\phi \right) \nabla
^{i}\phi -\left( D\phi \right) ^{2}\nabla _{b}D\phi +\left( D\phi \right)
^{2}\nabla _{b}\phi \left( \nabla _{i}l^{i}+D\phi \partial _{\phi }\right) %
\right] V~,
\end{equation}%
and%
\begin{equation}
T_{ab}^{(2)}=-l_{a}l_{b}D\phi \left( \nabla _{c}D\phi \right) \nabla
^{c}\phi V~,
\end{equation}%
necessary for the perturbative Kerr--Schild solution to be also exact,
through the second condition (\ref{cond21}) yield (unless $D\phi =0$)%
\begin{equation}
\left[ l_{(a}\nabla _{b)}D\phi -\left( \nabla _{c}l^{c}\right) l_{(a}\nabla
_{b)}\phi \right] V=l_{(a}\nabla _{b)}\phi D\phi V_{\phi }~.  \label{Vcond}
\end{equation}%
Either its trace or contracting with $l^{b}$ gives%
\begin{equation}
\left[ D^{2}\phi -\left( \nabla _{c}l^{c}\right) D\phi \right] V=\left(
D\phi \right) ^{2}V_{\phi }~.
\end{equation}%
Reinserting this in Eq. (\ref{Vcond}) results in%
\begin{equation}
\left[ l_{(a}\nabla _{b)}\phi D^{2}\phi -\left( l_{(a}\nabla _{b)}D\phi
\right) D\phi \right] V=0~.
\end{equation}%
With the exception of the very special scalar field obeying $l_{(a}\left(
\nabla _{b)}\phi D^{2}\phi -D\phi \nabla _{b)}D\phi \right) =0$, we obtain $%
V=0$. Hence Kerr--Schild type solutions of the linearized Einstein equation
sourced by pure kinetic gravity braiding and generated by autoparallel null
congruences belong to the 
\begin{equation}
H=0~.
\end{equation}%
class.

Note that the scalar equation for $H=-V\left( \phi \right) $ simplifies to 
\begin{align}
& 0=-V\left[ \left( \nabla _{a}\nabla _{b}\phi \right) \left( \nabla
^{a}\nabla ^{b}\phi \right) -\left( \square \phi \right) ^{2}\right]  \notag
\\
& +2V_{\phi }\left[ \nabla _{a}\phi \nabla _{b}\phi \left( \nabla ^{a}\nabla
^{b}\phi \right) -2X\square \phi \right] +4V_{\phi \phi }X^{2}  \notag \\
& \emph{+}2V^{2}\left( -2\nabla ^{a}\phi \nabla ^{b}\phi \nabla _{a}\nabla
_{b}\phi -X\square \phi \right) X+4VV_{\phi }X^{3}~.
\end{align}%
This is also solved for $V=0.$

With $B=0=V$ any $H$ linear in $X$ reduces to pure vacuum, as $H_{X}=B=0$
and $H_{\phi }=0$, therefore the scalar equation of motion becomes trivial.

\subsection{The $D\protect\phi =0$ case}

In the considerations above we excluded the $D\phi =0$ case. In this
subsection we consider this special scalar field.

With $D\phi =0$, Eqs. (\ref{nablaXKS}) and (\ref{dalKS}) imply%
\begin{equation}
\tilde{\nabla}_{c}\tilde{X}=\nabla _{c}X~,\quad \tilde{\square}\phi =\square
\phi ~.
\end{equation}%
The perturbative contributions (\ref{T1KGB}) and (\ref{TkKGB}) also simplify
as%
\begin{equation}
T_{ab}^{(1)}=l_{a}l_{b}\nabla _{c}X\nabla ^{c}\phi H~,\quad T_{ab}^{(k\geq
2)}=0~.
\end{equation}%
Next, we discuss the conditions for the linearized Einstein equation to
generate an exact Kerr--Schild spacetime. As $T_{ab}^{(3)}=0$, Eq. (\ref%
{cond3}) implies an autoparallel Kerr--Schild congruence. Further, the
second condition (\ref{autopcond}) is trivially satisfied. Therefore (as it
was the case for the k-essence), no restrictions on $H$ emerge in this case.
This is consequence of neither $\phi $ nor $X$ being affected by the
Kerr--Schild map in the special case of a scalar constant along the
congruence.

\section{Scalar constraint}

We have applied Kerr--Schild maps on the metric, which do not change the
scalar field. We equally assumed that the energy-momentum tensor of the
scalar has the same functional form before and after the map. In this
section we discuss, whether such assumptions could impose any constraints on
the scalar field. We illustrate this for the case (\ref{Flin}) of the
k-essence linear in $X$, for which the scalar equation (\ref{scalarEq})
becomes%
\begin{equation}
C\square \phi -U_{\phi }=C_{\phi }X~.
\end{equation}%
For $C=1$ this reproduces the Klein--Gordon type equation for quintessence
with potential $U$. Requiring the same equation to hold for the scalar after
the KS map:%
\begin{equation}
C\tilde{\square}\phi -U_{\phi }=C_{\phi }\tilde{X}~,
\end{equation}%
Eqs. (\ref{Xtil}) and (\ref{dalKS}) yield the constraint%
\begin{equation}
-2C\left[ D^{2}\phi +\left( \nabla _{a}l^{a}\right) D\phi \right] -U_{\phi
}=C_{\phi }\left( D\phi \right) ^{2}~.  \label{Sconstr}
\end{equation}%
This is a serious constraint on the possible scalar fields. Even in the
simplest case $C=1$, $U=0$ this gives%
\begin{equation}
D^{2}\phi =-\left( \nabla _{a}l^{a}\right) D\phi ~.
\end{equation}%
Such equations should be obeyed together with the linearized Kerr--Schild
maps when searching for perurbative solutions, which are also exact.

Note that in the special case $D\phi =0=U_{\phi }$,\ the constraint (\ref%
{Sconstr}) is trivially obeyed.

\section{Pp-waves in kinetic gravity braiding}

General relativity predicted gravitational waves, the existence of which was
spectacularly confirmed by more than 200 detections of gravitational waves
from coalescing compact binaries by LIGO-Virgo-KAGRA. In the geometrical
optics / high frequency approximation these waves are characterized by null
rays along which the waves propagate. This is similar to how electromagnetic
waves are presented in the geometrical optics approximation. The first
examples of gravitational waves were the family of cylindrically symmetric
Einstein--Rosen waves \cite{EinsteinRosen}, later shown to contain beyond
stationary solutions also solitonic and pulse-type solutions \cite{Delice}.
The simplest plane waves can be generalized into the class of plane-fronted
waves with parallel propagation (pp-waves) discussed below. A remarkable
result by Penrose states that any spacetime has a plane wave as a limit \cite%
{Penroselimit}.\ Plane waves are identical with their Penrose limit. Some of
the Penrose limit plane waves were shown to be diagonalizable \cite{Tod}.
The Penrose limit in general is a pp-wave \cite{Blau}, which, beside plane
waves also include the Aichelburg--Sexl ultraboost, an impulsive pp-wave
spacetime perceived by observers moving with high speed close to the speed
of light in the vicinity of a black hole \cite{AS}. The spacetime about null
geodesics can also be modelled through pp-waves.

\subsection{Pp-waves}

The line element of pp-waves in Brinkmann coordinates is manifestly in
Kerr--Schild form, Eq. (24.40) of Ref. \cite{Stephani} 
\begin{equation}
ds^{2}=\underbrace{-2dudv+dx^{2}+dy^{2}}_{\text{flat}}+L\left( u,x,y\right)
du^{2}~.
\end{equation}%
Whenever $L\left( u,x,y\right) $ is quadratic in $x$ and $y$, the spacetime
represents a plane wave with extra planar symmetry and the coordinates%
\begin{equation}
u=\frac{ct-z}{\sqrt{2}}~,\quad v=\frac{ct+z}{\sqrt{2}}
\end{equation}%
are null.

The $uu$-component of Ricci tensor is%
\begin{equation}
R_{uu}=-\frac{1}{2}\left( \partial _{x}^{2}+\partial _{y}^{2}\right) L\left(
u,x,y\right) ~,~
\end{equation}%
while the other components vanish. As $R=g^{uu}R_{uu}=0$, the Einstein
equations read $R_{ab}=\left( c^{4}/8\pi G\right) T_{ab}$.

\subsection{K-essence pp-waves}

Pp-waves emerging from a k-essence source with linear dependence on $X$ fall
into the $F=0$ class. (This is, because the $F=0$ class is not necessarely
empty, similarly to $R=0$ containing all vacuum metrics). Indeed, the $vv$
component of the Einstein equation leads to $\partial _{v}\phi =0$, thus $%
\phi =\phi \left( u,x,y\right) $. Then the $uv$ component immediately gives $%
F=0$.

Inserting this into the $xx$, $yy$ components yields either $C=0=U$ (the
latter stemming then from $F=0$), thus vanishing energy-momentum or for $%
C\neq 0$ 
\begin{equation}
\partial _{x}\phi =0~,\quad \partial _{y}\phi =0~,
\end{equation}%
thus $\phi =\phi \left( u\right) $, a harmonic scalar field. The $uu$
component of the Einstein equation remains the only nontrivial one:%
\begin{equation}
\left( \partial _{x}^{2}+\partial _{y}^{2}\right) L\left( u,x,y\right) =-%
\frac{c^{4}}{4\pi G}C\left( \partial _{u}\phi \right) ^{2}~.
\label{F0kessence}
\end{equation}%
In the special case of constant $\phi $ this reproduces the vacuum pp-waves 
\cite{Ehlers}, with solutions including all functions 
\begin{equation}
L_{\text{hom}}=L_{\text{hom}}\left( u\right)  \label{Lhom}
\end{equation}%
and other, $x$ and $y$ dependent solutions possible for particular boundary
conditions. A particular solution of the inhomogeneous equation (\ref%
{F0kessence}) emerges as follows. The solution of the two-dimensional
Poisson equation $\left( \partial _{x}^{2}+\partial _{y}^{2}\right) P\left(
x,y\right) =1$ is $P=\left( x^{2}+y^{2}\right) /4$, therefore%
\begin{equation}
L_{\text{inhom}}\left( u,x,y\right) =-\frac{c^{4}}{16\pi G}C\left( \partial
_{u}\phi \right) ^{2}\left( x^{2}+y^{2}\right) ~,  \label{Linhom}
\end{equation}%
represents a plane wave (being quadratic in $x$ and $y$). Thus, the general
solution is 
\begin{equation}
L=L_{\text{hom}}\left( u\right) +L_{\text{inhom}}\left( u,x,y\right) ~.
\label{Lgen}
\end{equation}%
Note that for $C\neq 0$ the condition $F=0$ could be obeyed only with 
\begin{equation}
U=0=X~.  \label{condppwave}
\end{equation}%
The latter condition means either vacuum or that the gradient of the scalar
field is a null vector. Canonical scalar fields with vanishing potential and
null gradient were shown to represent expansionless null dust \cite%
{Faraoni2018,Faraoni}.

\subsection{\textbf{The scalar constraint is satisfied}}

The Kerr--Schild vector field

\begin{equation}
l^{a}\partial _{a}=-\left\vert L\left( u,x,y\right) \right\vert
^{1/2}\partial _{v}~
\end{equation}%
and $\partial _{v}\phi =0$ (obtained from the Einstein equations) allows to
show that

\begin{equation}
D\phi =l^{a}\nabla _{a}\phi =l^{v}\partial _{v}\phi =0~.
\end{equation}%
With (\ref{condppwave}) holding the scalar constraint (\ref{Sconstr}) is
identically satisfied.

\subsection{The tower of Kerr--Schild maps}

The first such map takes the flat seed metric 
\begin{equation}
ds^{2}=\underbrace{-2dudv+dx^{2}+dy^{2}}_{\text{flat}}~
\end{equation}%
into a vacuum pp-wave with $L_{\text{hom}}$:%
\begin{equation}
d\tilde{s}^{2}=\underbrace{-2dudv+dx^{2}+dy^{2}}_{\text{flat}}+L_{\text{hom}%
}\left( u\right) du^{2}~,
\end{equation}%
with the null Kerr--Schild congruence $l_{a}dx^{a}=\left\vert L_{\text{hom}%
}\left( u\right) \right\vert ^{1/2}du$ (this is in the absence of $\phi $).

A subsequent Kerr--Schild map with $l_{a}dx^{a}=\left\vert L_{\text{inhom}%
}\left( u,x,y\right) \right\vert ^{1/2}du$ then takes $d\tilde{s}^{2}$ into%
\begin{equation}
d\widetilde{\tilde{s}}^{2}=\underbrace{-2dudv+dx^{2}+dy^{2}+L_{\text{hom}%
}\left( u\right) du^{2}}_{\text{vacuum pp-wave with }L_{\text{hom}}\text{
(in the absence of }\phi \text{)}}+L_{\text{inhom}}\left( u,x,y\right)
du^{2}~,
\end{equation}%
another pp-wave with $L=L_{\text{hom}}+L_{\text{inhom}}$.

Application of further Kerr-Schild maps could follow, changing the retarded
time dependent function $C\left( \phi \right) $ in $L_{\text{inhom}}$.

This construction is very similar to Kerr--Schild maps generating
Schwarzschild from flat spacetime, followed by other Kerr--Schild maps
merely changing the mass.

\section{Conluding remarks}

Exact vacuum Kerr--Schild spacetimes can be recovered as solutions of the
linearized Einstein equations. This advantageous property was shown to
persist in the presence of specific matter sources. K-essence scalar fields
could source exact Kerr--Schild spacetimes induced by the perturbative
solution for Lagrangian either linear in the kinetic term (in this case the
Kerr--Schild congruence being autoparallel) or unrestricted, when the scalar
is constant along the Kerr--Schild congruence, $D\phi =0$. In this paper we
generalized the source term to include the full class of kinetic braiding
scalar fields. We proved that the property withstands for the pure kinetic
gravity braiding contribution only for a vanishing Lagrangian (equivalent to
vacuum for this case) or in the case $D\phi =0$ (a scalar constant along the
Kerr--Schild congruence). We also showed that the requirement of an
unchanged scalar field and an unchanged functional form of its
energy-momentum tensor under the Kerr--Schild map induces a scalar
constraint, which has to be considered in addition to the linearized
Kerr--Schild equation.

Finally, we discussed pp-waves, which are manifestly of Kerr--Schild type.
Beyond vacuum pp-waves we also derived those which are generated by a
k-essence with vanishing Lagrangian and $D\phi =0$ (such that they obey the
scalar constraint). At the end, we identified a Fock space type construction
through successive application of a tower of Kerr--Schild maps.

Gravitational waves in the geometrical optics approximation are pp-waves.
With the forthcoming LISA and other prospective space detectors, able to
monitor gravitational wave for weeks or even months, vacuum pp-waves (with $%
C=0$) could in principle be distinguished from those sourced by k-essence
(with $C\neq 0$), raising the possibility to constraint such scalar fields.
Working out such methods would provide another challenging way for testing
general relativity.


\begin{thebibliography}{99}
\bibitem{KerrSchild} Kerr K.P.; Schild. A. A new class of vacuum solutions
of the Einstein field equations, Atti del convegno sulla relativit`a
generale; problemi dellenergia e onde gravitationali, IV Centenario Della
Nascita di Galileo Galilei, Barb\`{e}ra Editore, Firenze, Italy; \textbf{%
1965; }Volumen 2,\textbf{\ }222.

\bibitem{GALPZ} Gergely, L.\'{A}.; Perj\'{e}s, Z. Kerr-Schild metrics
revisited I. The ground state. J. Math. Phys. \textbf{1994}, 35, 2438.

\bibitem{GALPZ2} Gergely, L.\'{A}.; Perj\'{e}s, Z. Kerr-Schild metrics
revisited II. The complete vacuum solution. J. Math. Phys. \textbf{1994},
35, 2448.

\bibitem{GALPZ3} Gergely, L.\'{A}.; Perj\'{e}s, Z. Vacuum Kerr-Schild
metrics generated by nontwisting congruences. Ann. Phys. \textbf{1994}, 3,
609.

\bibitem{KentZimmerman} Kent, B.; Zimmerman, A. A new framework for
classical double copies, \textbf{2025}, arXiv:2505.03887 [hep-th].

\bibitem{Xanthopoulos} Xanthopoulos, B.C. Exact vacuum solutions of
Einstein's equation from linearized solutions. J. Math. Phys. \textbf{1978},
19, 1607.

\bibitem{GAL} Gergely, L. \'{A}. Linear Einstein equations and Kerr-Schild
maps, Class. Quant. Grav. \textbf{2002,} 19\textbf{,} 2515.

\bibitem{KGB} Deffayet, C.; Pujolas, O.; Sawicki, I.; Vikman, A. Imperfect
Dark Energy from Kinetic Gravity Braiding. JCAP \textbf{2010}, 1010, 026.

\bibitem{Horndeski1} Horndeski, G.W. Second-order scalar-tensor field
equations in a four-dimensional space. Int. J. Theor. \textbf{1974}, 10,
363--384.

\bibitem{Deffeayet} Deffayet, C.; Gao, X.; Steer, D.A.; Zahariade, G. From
k-essence to generalized Galileons. Phys. Rev. D. \textbf{2011}, 84, 064039.

\bibitem{GW} Abbott, B.P. et al. [LIGO Scientific and Virgo Collaborations,
Fermi Gamma-ray burst monitor, and INTEGRAL] Gravitational Waves and
Gamma-Rays from a Binary Neutron Star Merger: GW170817 and GRB170817A.
Astrophys. J. Lett. \textbf{2017}, 848, L13.

\bibitem{GW0} Baker, T.; Bellini, E.; Ferreira, P.G.; Lagos, M.; Noller, J.;
Sawicki, I. Strong constraints on cosmological gravity from GW170817 and GRB
170817A. Phys. Rev. Lett. \textbf{2017}, 119, 251301.

\bibitem{GW1} Ezquiaga, J.M.; Zumalac\'{a}rregui, M. Dark Energy after
GW170817: Dead ends and the road ahead. Phys. Rev. Lett. \textbf{2017}, 119,
251304.

\bibitem{GW2} Creminelli, P.; Vernizzi, F. Dark Energy after GW170817 and
GRB170817A. Phys. Rev. Lett. \textbf{2017}, 119, 251302.

\bibitem{GW3} Sakstein, J.; Jain, B. Implications of the Neutron Star Merger
GW170817 for Cosmological Scalar-Tensor Theories. Phys. Rev. Lett. \textbf{%
2017}, 119, 251303.

\bibitem{bigrip} Vasilev, T. B.; Bouhmadi-L\'{o}pez, M.; Mart\'{\i}n-Moruno,
P. Big rip in shift-symmetric Kinetic Gravity Braiding theories, Phys. Lett.
B \textbf{2023,} 838, 137711.

\bibitem{phantom} Vasilev, T. B.; Bouhmadi-L\'{o}pez, M.; Mart\'{\i}%
n-Moruno, P. Phantom attractors in Kinetic Gravity Braiding theories: a
dynamical system approach, JCAP \textbf{2023}, 06, 026.

\bibitem{kessence} Juh\'{a}sz, B.; Gergely, L.\'{A}. K-Essence Sources of
Kerr--Schild Spacetimes, Universe \textbf{2025}, 11(3), 100.

\bibitem{KGBFluid} Gergely, L. \'{A}. Fluid interpretations of the scalar
field in kinetic gravity braiding, \textbf{2025}.

\bibitem{EinsteinRosen} A. Einstein, N. Rosen, On gravitational waves. J.
Franklin Inst. \textbf{1937}, 223, 43.

\bibitem{Delice} Akyar L.; Delice \"{O}. On generalized Einstein-Rosen waves
in Brans-Dicke theory, Eur. Phys. J. Plus \textbf{2014,} 129, 226.

\bibitem{Penroselimit} Penrose R. Any space-time has a plane-wave as a
limit, in Differential geometry and relativity, Eds. Cahen M.; Flato M., 
\textbf{1976,} 271.

\bibitem{Tod} Tod P. Spacetimes with all Penrose limits diagonalisable
Class.Quant.Grav. \textbf{2020}, 37, 7, 075021.

\bibitem{Blau} Blau M. Plane waves and Penrose Limits, \textbf{2024},
http://blau.itp.unibe.ch/lecturesPP.pdf

\bibitem{AS} Aichelburg, P. C.; Sexl, R. U. On the gravitational field of a
massless particle, General Relativity and Gravitation, \textbf{1971}, 2(4),
303.

\bibitem{Stephani} Stephani, H.; Kramer, D.; Maccallum, M.; Hoenselaers, C.;
Herlt, E. Exact Solutions of Einstein's Field Equations; Cambridge
University Press: Cambridge, UK, \textbf{2023}; Volume 485.

\bibitem{Ehlers} Ehlers J.; Schmidt B. G. Einstein's Field Equations and
Their Physical Implications: Selected Essays in Honour of J\"{u}rgen Ehlers,
Springer-Verlag New York, LLC, \textbf{2000} Lecture notes in physics 540.

\bibitem{Faraoni2018} Faraoni, V.; C\^{o}t\'{e}, J. Scalar field as a null
dust, Eur. Phys. J. C \textbf{2019}, 79, 318.

\bibitem{Faraoni} Faraoni, V.; Giusti, A.; H. Fahim, B. H. Vaidya geometries
and scalar fields with null gradients, Eur. Phys. J. C \textbf{2021, }81,
232.
\end{thebibliography}
\end{document}